\documentclass[a4paper]{article}
\usepackage[utf8]{inputenc}
\usepackage[T1]{fontenc}
\usepackage{graphicx}
\usepackage{grffile}
\usepackage{longtable}
\usepackage{wrapfig}
\usepackage{rotating}
\usepackage[normalem]{ulem}
\usepackage{amsmath}
\usepackage{textcomp}
\usepackage{amssymb}
\usepackage{capt-of}
\usepackage{hyperref}
\usepackage{pdfpages}
\usepackage{graphicx}
\usepackage{amssymb}
\usepackage{booktabs}
\usepackage{xcolor}
\usepackage[utf8]{inputenc}
\usepackage{multirow}
\usepackage{caption}
\usepackage[margin=2cm]{geometry}
\usepackage{fancyhdr}
\usepackage{sourcecodepro}
\usepackage{array}
\usepackage{colortbl}
\usepackage{listings}
\usepackage[english]{babel}
\usepackage[scale=2]{ccicons}
\usepackage{hyperref}
\hypersetup{hidelinks=true}
\usepackage{relsize}
\usepackage{amsmath}
\usepackage{bm}
\usepackage{amsfonts}
\usepackage{bm}
\usepackage{wasysym}
\usepackage{float}
\usepackage{ragged2e}
\usepackage{textcomp}
\usepackage{pgfplots}
\usepackage{todonotes}
\usepgfplotslibrary{dateplot}
\definecolor{Accent}{HTML}{157FFF}

\graphicspath{{./img/}}

\usepackage{RR}
\RRNo{9310}
\RRdate{December 2019}
\RRauthor{Tom Cornebize \and Arnaud Legrand}
\authorhead{Cornebize \& Legrand}
\RRetitle{DGEMM performance is data-dependent}
\RRtitle{Les performances de DGEMM dépendent des données}
\RRabstract{The DGEMM function is a widely used implementation of the matrix product. While the asymptotic complexity of the algorithm only depends on the sizes of the matrices, we show that the performance is significantly impacted by the matrices content. Our experiments show that this may be due to bit flips in the CPU causing an energy consumption overhead.}
\RRresume{La fonction DGEMM est une implémentation très utilisée du produit matriciel. Bien que la complexité asymptotique de l'algorithme ne dépende que des tailles des matrices, nous montrons que les performances sont significativement impactées par le contenu des matrices. Nos expériences montrent que cela pourrait être causé par une surconsommation énergétique due aux changements d'état dans le processeur.}
\RRkeyword{performance,DVFS,BLAS,DGEMM}
\RRmotcle{performance,DVFS,BLAS,DGEMM}
\URRhoneAlpes
\RRprojets{POLARIS}
\date{}
\title{DGEMM performance is data-dependent}
\hypersetup{
 pdfauthor={Tom Cornebize, Arnaud Legrand},
 pdftitle={DGEMM performance is data-dependent},
 pdfkeywords={},
 pdfsubject={},
 pdfcreator={Emacs 26.1 (Org mode 9.1.13)}, 
 pdflang={English}}
\begin{document}

\makeRR
\section{Context}
\label{sec:org9d5b8cb}
In a recent work, we have shown how to predict the performance of an MPI
application efficiently and faithfully\cite{smpi_hpl}, using High Performance
Linpack (HPL) as a case study. We needed to build a performance model for
several computation kernels, including \texttt{dgemm}. This was done by running these
kernels on all the nodes of the platform with various parameter combinations and
measuring their durations. We noticed a significant mismatch between the
durations measured with our calibration code and the durations observed in
HPL. We found out that, the performance of the \texttt{dgemm} function depends on the
content of the matrix, which was unexpected.

The different experiments presented in this report have been repeated on several
nodes and follow the same steps:
\begin{enumerate}
\item Deploy and install a fresh OS on the node.
\item Run the \texttt{stress} command for 10 minutes to warm the node.
\item Start a background process\footnote{\url{https://github.com/Ezibenroc/Stress-Test/blob/8a33d1f/basic\_monitoring.py}} to monitor the core frequencies every
second.
\item On each core, run a custom code\footnote{\url{https://github.com/Ezibenroc/platform-calibration/blob/153a4a1/src/calibration/calibrate\_blas.c}} that will perform a sequence of calls
to the \texttt{dgemm} function with a single thread and three square matrices of order
2,048. The durations of each \texttt{dgemm} call is written in a CSV file. The only
change in the different experiments we did is how the three matrices are
initialized.
\end{enumerate}

The plots shown in this report have been obtained on four nodes of the Dahu
cluster from Grid'5000\footnote{\url{https://www.grid5000.fr/w/Grenoble:Hardware}}. Each of these nodes has two Intel Xeon Gold 6130
CPU, which are 16 core CPU from the Skylake generation. They have a base
frequency of 2.1 GHz and a turbo frequency of up to 3.7 GHz, but their turbo
frequency is limited to 2.4 GHz when their 16 cores are active and in AVX2
mode\footnote{\url{https://en.wikichip.org/wiki/intel/xeon\_gold/6130}}. We have used OpenBLAS\footnote{\url{https://github.com/xianyi/OpenBLAS}} version 0.3.1, compiled with GCC version
6.3.0 on a Debian 9 installation with kernel version 4.9.0.

Note that one of the nodes we used, dahu-15, is known to have severe cooling
issues that lead to significant frequency and performance drops\footnote{\url{https://intranet.grid5000.fr/bugzilla/show\_bug.cgi?id=10270}}.
\section{Randomization of the matrix initialization}
\label{sec:orgf430fbc}
The three matrices are allocated once at the start of the program as a buffer of
size \(N^2\) with \(N=2,048\). Then, their content is initialized in three different
ways, depending on the experiment:
\begin{enumerate}
\item All the elements of the matrices are equal to some constant. We have tested
with three different values: 0, 0.987 and 1.
\item The elements of the matrices are made of an increasing sequence in the
interval \([0, 1]\). More precisely, \texttt{mat[i] = i/(N\textasciicircum{}2-1)} for i in \([0, N^2-1]\).
\item Each element of the matrix is randomly and uniformly sampled in the interval
\([0, 1]\).
\end{enumerate}

\begin{figure}[htbp]
\centering
\includegraphics[width=0.9\textwidth]{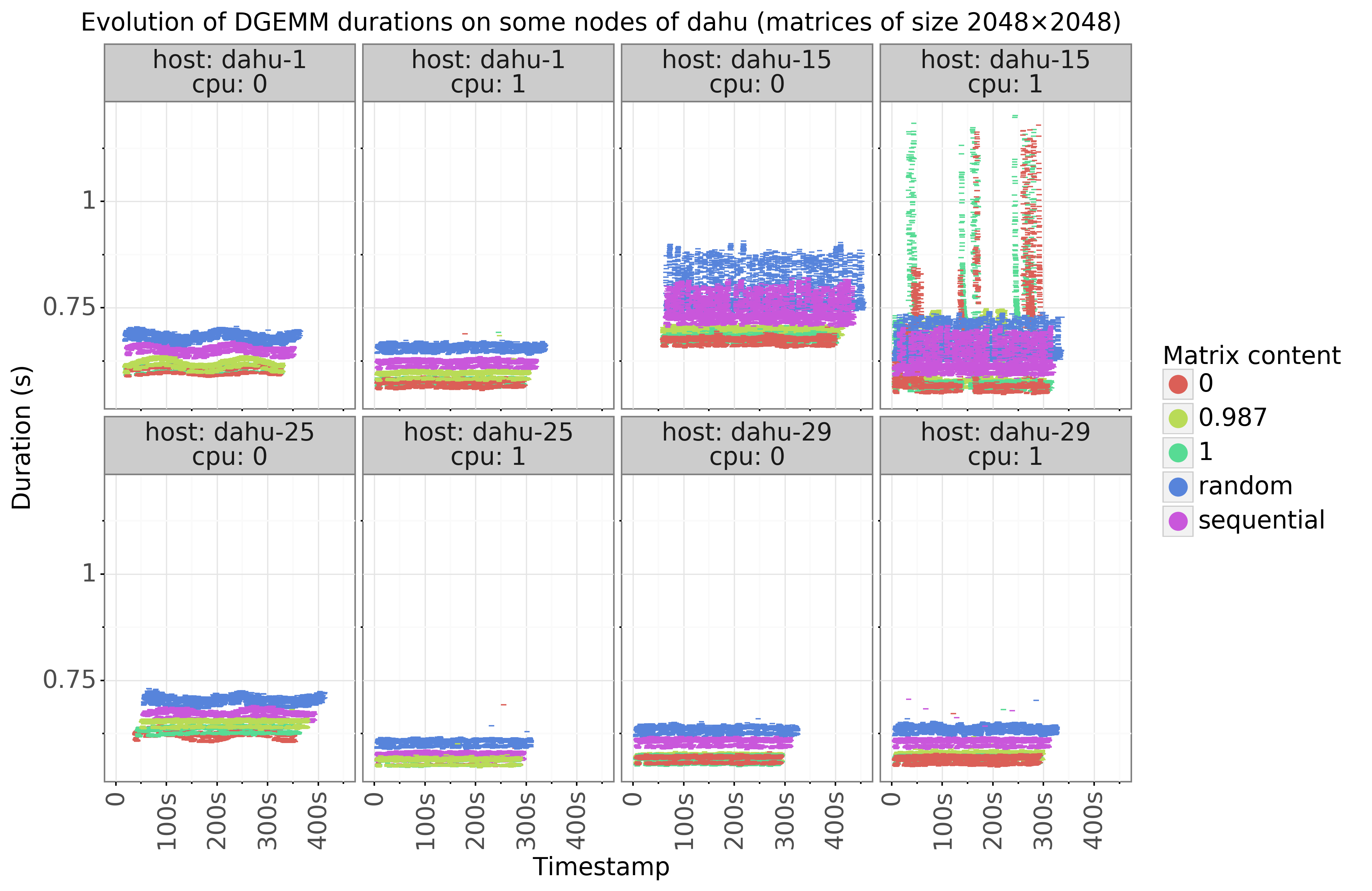}
\caption{\label{fig:org4ac2157}
DGEMM durations are lower with constant values in the matrices}
\end{figure}

Figure \ref{fig:org4ac2157} shows the evolution of the dgemm
durations during the experiment. Several temporal patterns can be distinguished:
the CPU 0 of dahu-1 and dahu-25 have oscillating performance, the CPU 1 of
dahu-15 has huge performance drops. More interestingly, several layers can be
seen: the random points seem to be on top of the sequential points which are
above the constant ones.

\begin{figure}[htbp]
\centering
\includegraphics[width=0.9\textwidth]{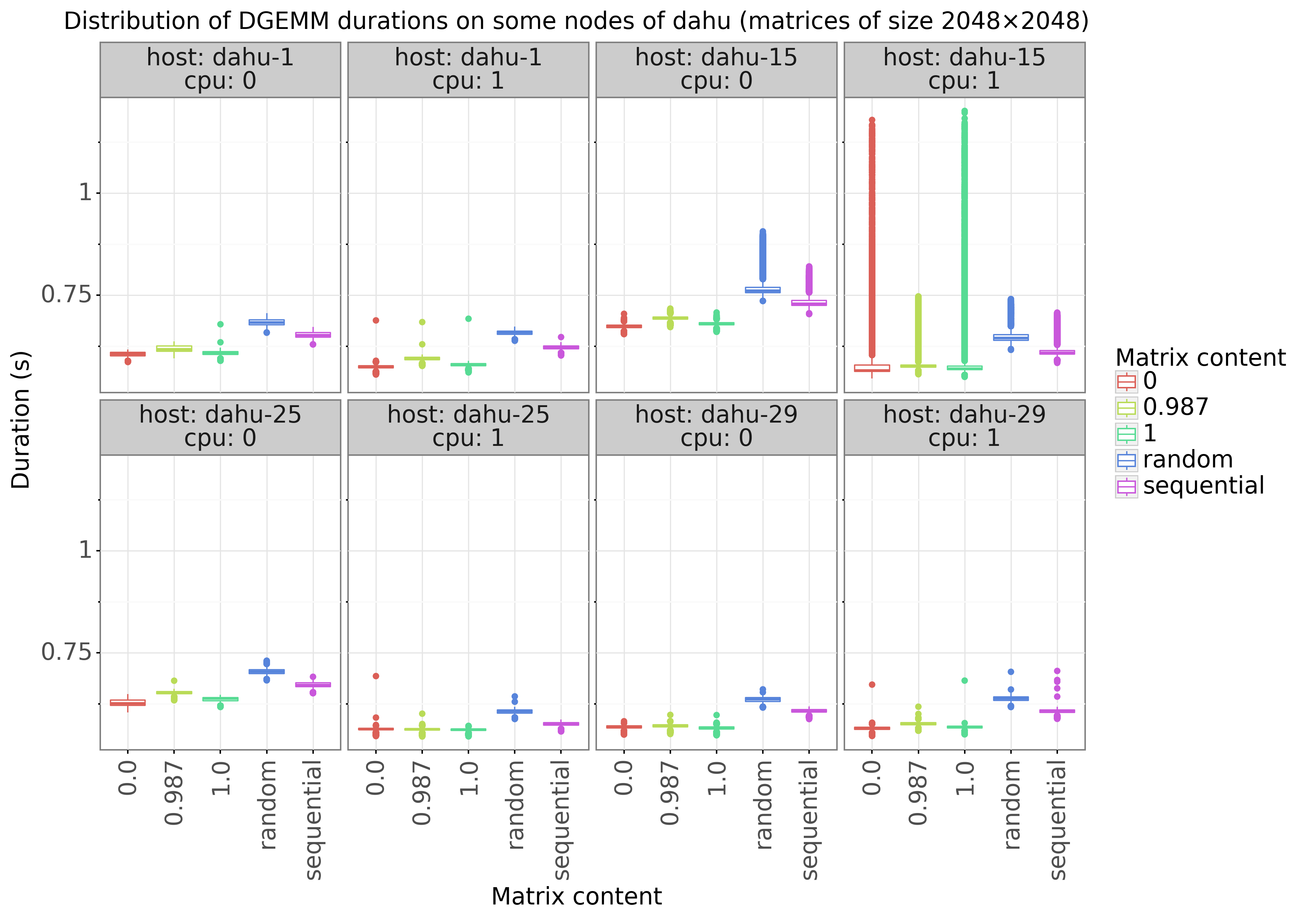}
\caption{\label{fig:org119f32d}
DGEMM durations are lower with constant values in the matrices}
\end{figure}

This last observation is confirmed by the box plots shown in figure
\ref{fig:org119f32d}. We see here that on the four nodes, the
durations of \texttt{dgemm} are the highest when the matrices are initialized randomly
and the lowest when they are initialized with a constant value. The sequential
initialization is in between.

Such an observation was unforeseen. The function \texttt{dgemm} implements the usual
matrix product with cubic complexity. The control flow of the function does not
depend on the matrix content, so we did not expect its duration to be
data-dependent.

\begin{figure}[htbp]
\centering
\includegraphics[width=0.9\textwidth]{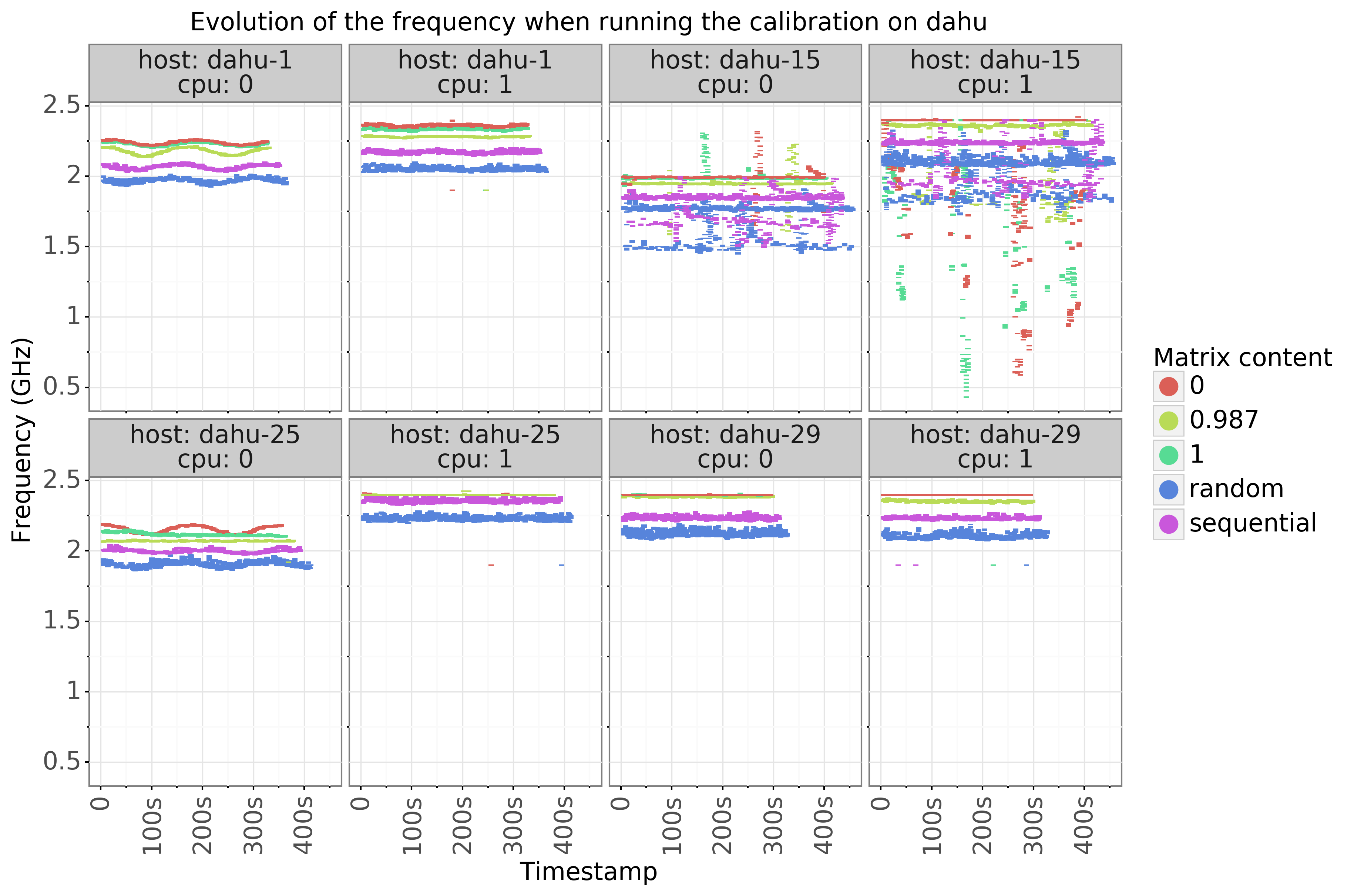}
\caption{\label{fig:org52b93a8}
Core frequencies are higher with constant values in the matrices}
\end{figure}

\begin{figure}[htbp]
\centering
\includegraphics[width=0.9\textwidth]{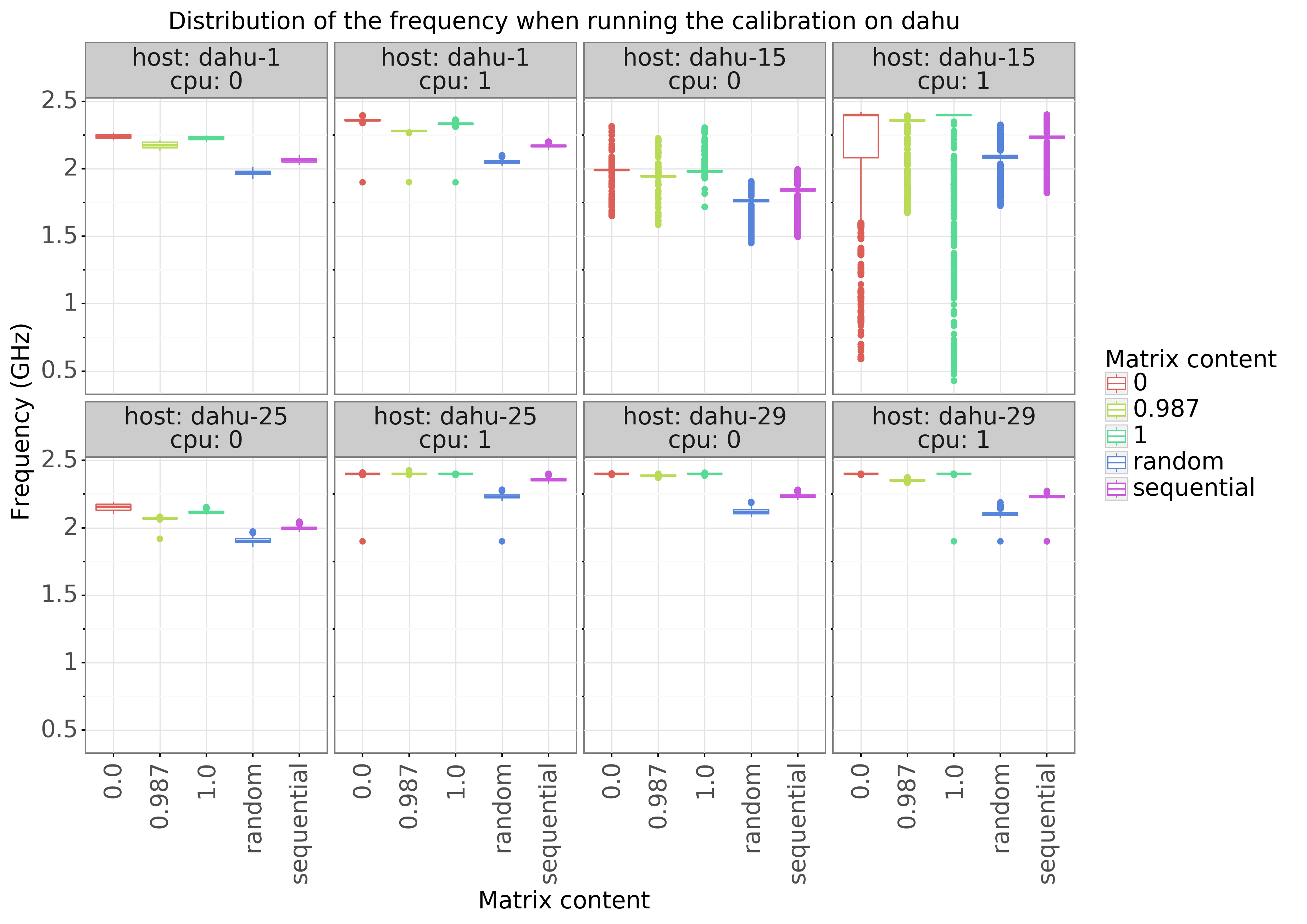}
\caption{\label{fig:orgb6f5106}
Core frequencies are higher with constant values in the matrices}
\end{figure}

The observations we have made on \texttt{dgemm} performance can be explained by figures
\ref{fig:org52b93a8} and \ref{fig:orgb6f5106} which
respectively show the evolution of the core frequencies over time and the
distributions of these frequencies. There is a clear correlation between the
frequencies and \texttt{dgemm} performance: the random initialization produces lower
frequencies whereas the constant initialization gives higher
frequencies. Similar temporal patterns can also be distinguished: the
oscillation on the CPU 0 of dahu-1 and dahu-25 as well as the huge drops for the
CPU 1 of dahu-15.

This experiment has been repeated on other Grid'5000 clusters. Table
\ref{tab:org6b02740} gives a summary of our observations. Four other clusters show a
similar behavior, the performance of \texttt{dgemm} is higher when the matrices are
generated with a constant value. However, for four other clusters, this
phenomenon could not be observed, the matrix content had no impact on the
performance.

\begin{table}[htbp]
\caption{\label{tab:org6b02740}
Observation of the  performance anomaly on Grid'5000 clusters}
\centering
\begin{tabular}{llll}
\toprule
Cluster & CPU & Generation & Performance anomaly\\
\midrule
dahu & Intel Xeon Gold 6130 & Skylake & yes\\
yeti & Intel Xeon Gold 6130 & Skylake & yes\\
ecotype & Intel Xeon E5-2630L v4 & Broadwell & yes\\
gros & Intel Xeon Gold 5220 & Cascade Lake & yes\\
parasilo & Intel Xeon E5-2630 v3 & Haswell & yes\\
paranoia & Intel Xeon E5-2660 v2 & Ivy Bridge & no\\
nova & Intel Xeon E5-2620 v4 & Broadwell & no\\
taurus & Intel Xeon E5-2630 & Sandy Bridge & no\\
chiclet & AMD EPYC 7301 & - & no\\
\bottomrule
\end{tabular}
\end{table}
\section{Hypotheses}
\label{sec:orgf02665c}
Several hypotheses were discussed to explain this unexpected phenomenon.

There could be a small cache  on the floating point unit of the cores to
memorize the results of frequent operations. This could explain why the
durations were higher when the matrices were initialized randomly, but this does
not explain why the sequential initialization is in between.

This could be due to kernel same page merging (KSM), a mechanism that allows the
kernel to share identical memory pages between different processes. Again, this
would explain the difference between the random initialization and the constant
one, but not why the sequential initialization gives intermediate performance.

A last hypothesis is the power consumption of the cores. Each state change of
the electronic gates of the CPU costs an energy overhead. In the
case of the constant initialization, the registers will change less often during
the execution of \texttt{dgemm}, in comparison with the random initialization. As for the
sequential initialization, we can imagine that we have a locality effect: nearby
elements of the matrices will have more bits in common, this would causes less
bit flips than the random initialization but more bit flips than the constant
initialization and thus an intermediate performance.
\section{Testing the bit-flip hypothesis}
\label{sec:org0cd981d}
To test the hypothesis that the lower frequencies are caused by more frequent
bit flips in the processor, the matrix initialization has been changed. Now,
each element of the matrix is randomly and uniformly sampled in the interval
\([0,1]\). Then a bit mask is applied on the lower order bits of their
mantissa. As a result, all the elements of the matrices have some bits in
common. Several mask sizes have been tested, from 0 (the elements are left
unchanged) to 53 (the mantissa becomes completely deterministic, all the elements
are equal).

\begin{figure}[htbp]
\centering
\includegraphics[width=0.9\textwidth]{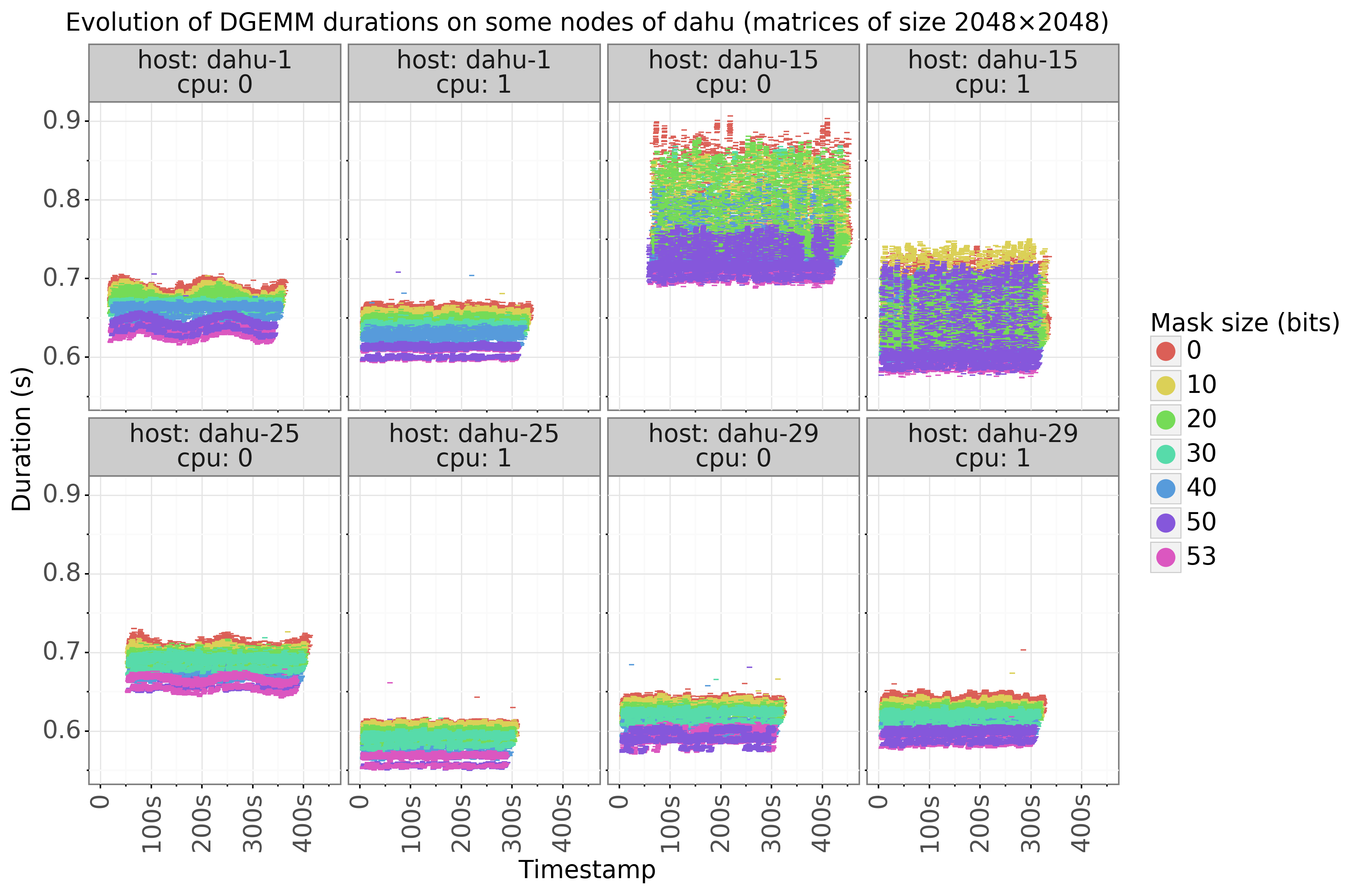}
\caption{\label{fig:org51f8c7a}
DGEMM durations are lower with larger bit masks}
\end{figure}

\begin{figure}[htbp]
\centering
\includegraphics[width=0.9\textwidth]{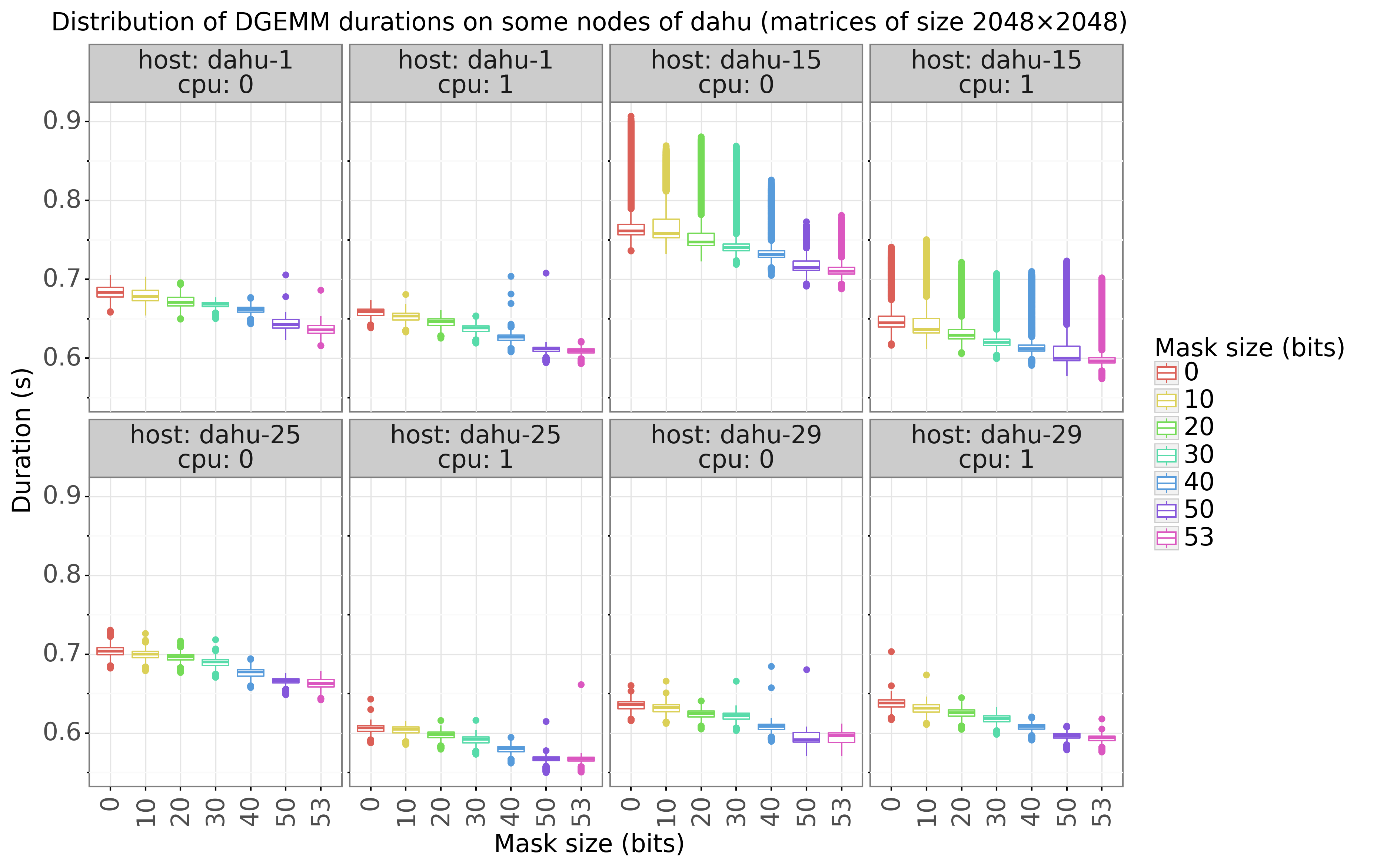}
\caption{\label{fig:orgd8602be}
DGEMM durations are lower with larger bit masks}
\end{figure}

The evolution of the dgemm durations is plotted in figure
\ref{fig:org51f8c7a}. Their distribution is shown in figure
\ref{fig:orgd8602be}. Their is a very clear correlation between the
mask size and the performance: the larger the mask, the lower the
duration. Similarly to the previous experiment, some temporal patterns can also
be distinguished.

\begin{figure}[htbp]
\centering
\includegraphics[width=0.9\textwidth]{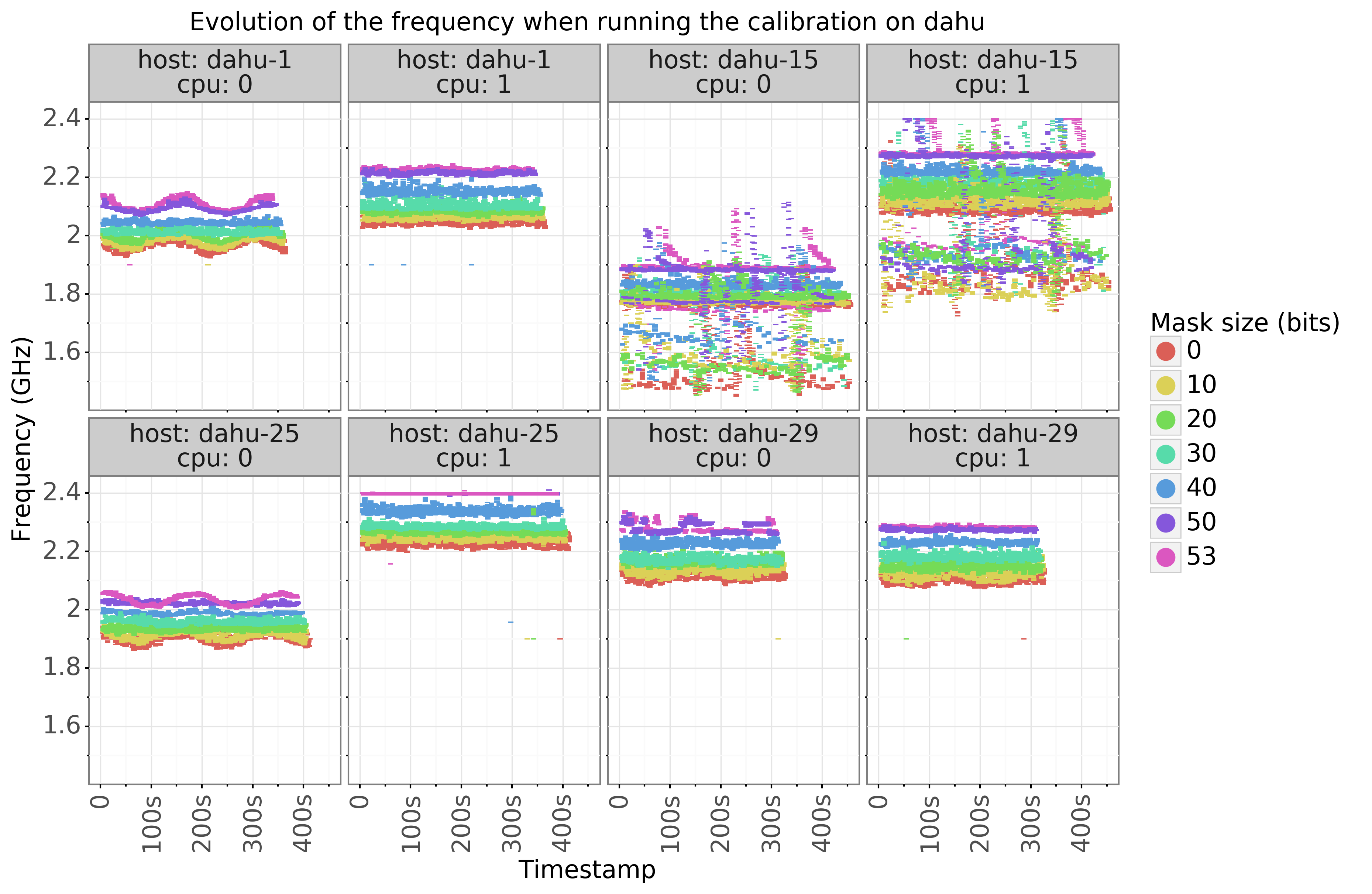}
\caption{\label{fig:orgf13c838}
Core frequencies are higher with larger bit masks}
\end{figure}

\begin{figure}[htbp]
\centering
\includegraphics[width=0.9\textwidth]{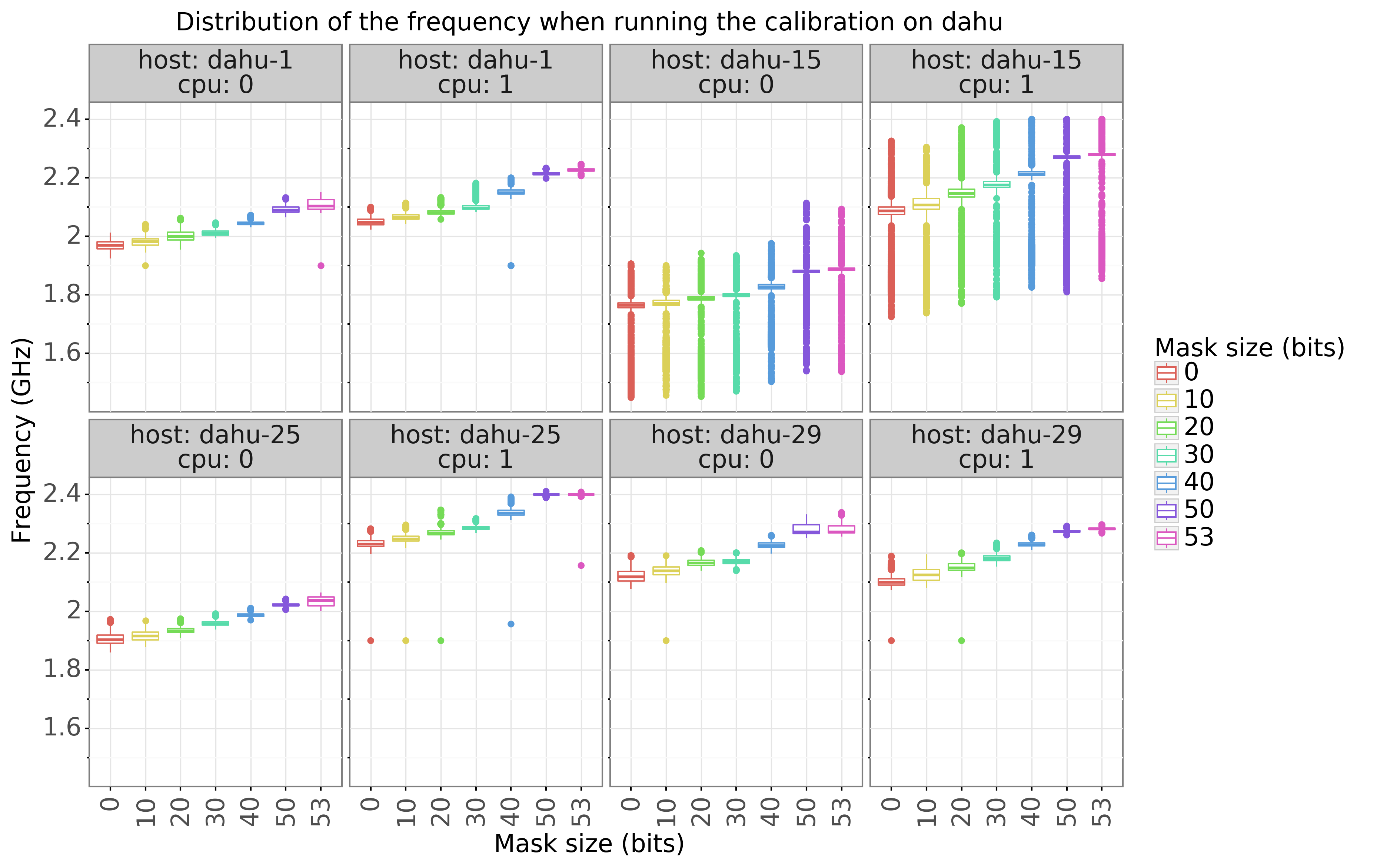}
\caption{\label{fig:org805e7b5}
Core frequencies are higher with larger bit masks}
\end{figure}

This correlation with the mask size can also be seen with the frequencies in
figures \ref{fig:orgf13c838} and \ref{fig:org805e7b5}: larger
masks lead to higher frequencies.

This experiment has been repeated on two other Grid'5000 clusters, ecotype and
gros. For both of them, the same observations could be made, a clear correlation
between the mask size, the frequencies and the performance.
\section{Conclusion}
\label{sec:org3ba4121}
We have shown that the performance of the \texttt{dgemm} function is data-dependent. The
best explanation we have for this counter-intuitive fact is an energy cost
overhead caused by bit flips inside the processor. This theory has been
corroborated by a controlled experiment where the elements of the matrices are
initialized semi-randomly: they all share an identical bit suffix.

To strengthen this claim further, the next steps will be to perform a similar
experiment with another compiler, another BLAS library and/or another
computation kernel. We also need to understand why some processors are subject
to this phenomenon and some others are not. Last, we wonder whether this could
be applied to perform a side-channel attack.
\section{Acknowledgements}
\label{sec:org206dcf2}
We warmly thank our colleagues that helped us to find hypotheses for this
performance anomaly. In particular, Guillaume Huard suggested that the
performance anomaly may be caused by the bit flips in the processor.

Experiments presented in this paper were carried out using the Grid'5000
testbed, supported by a scientific interest group hosted by Inria and including
CNRS, RENATER and several Universities as well as other organizations (see
\url{https://www.grid5000.fr}).
\bibliographystyle{IEEEtran}
\bibliography{refs}

\begin{thebibliography}{1}
\providecommand{\url}[1]{#1}
\csname url@samestyle\endcsname
\providecommand{\newblock}{\relax}
\providecommand{\bibinfo}[2]{#2}
\providecommand{\BIBentrySTDinterwordspacing}{\spaceskip=0pt\relax}
\providecommand{\BIBentryALTinterwordstretchfactor}{4}
\providecommand{\BIBentryALTinterwordspacing}{\spaceskip=\fontdimen2\font plus
\BIBentryALTinterwordstretchfactor\fontdimen3\font minus
  \fontdimen4\font\relax}
\providecommand{\BIBforeignlanguage}[2]{{%
\expandafter\ifx\csname l@#1\endcsname\relax
\typeout{** WARNING: IEEEtran.bst: No hyphenation pattern has been}%
\typeout{** loaded for the language `#1'. Using the pattern for}%
\typeout{** the default language instead.}%
\else
\language=\csname l@#1\endcsname
\fi
#2}}
\providecommand{\BIBdecl}{\relax}
\BIBdecl

\bibitem{smpi_hpl}
\BIBentryALTinterwordspacing
T.~Cornebize, A.~Legrand, and F.~C. Heinrich, ``{Fast and Faithful Performance
  Prediction of MPI Applications: the HPL Case Study},'' in \emph{{2019 IEEE
  International Conference on Cluster Computing (CLUSTER)}}, ser. 2019 IEEE
  International Conference on Cluster Computing (CLUSTER), Albuquerque, United
  States, Sep. 2019. [Online]. Available:
  \url{https://hal.inria.fr/hal-02096571}
\BIBentrySTDinterwordspacing

\end{thebibliography}
\end{document}